# THE INTERPLAY OF EXTINCTION AND SYNCHRONY IN THE DYNAMICS OF METAPOPULATION FORMATION

John Vandermeer and Zachary Hajian-Forooshani

Department of Ecology and Evolutionary Biology
University of Michigan
Ann Arbor, MI 48109

Corresponding author. J. Vandermeer
jvander@umich.edu

**Abstract**

**The idea of a metapopulation has become canonical in ecology. Its original mean field form provides the important intuition that migration and extinction interact to determine the dynamics of a population composed of subpopulations. From its conception, it has been evident that the very essence of the metapopulation paradigm centers on the process of local extinction. We note that there are two qualitatively distinct types of extinction, gradual and catastrophic, and explore their impact on the dynamics of metapopulation formation using discrete iterative maps. First, by modifying the classic logistic map with the addition of the Allee effect, we show that catastrophic local extinctions in subpopulations are a pre-requisite of metapopulation formation. When subpopulations experience gradual extinction, increased migration rates force synchrony and drive the metapopulation below the Allee point resulting in migration induced destabilization of the system across parameter space. Second, a sawtooth map (an extension of the Bernoulli bit shift map) is employed to simultaneously explore the increasing and decreasing modes of population behavior. We conclude with four generalizations. 1. At low migration rates, a metapopulation may go extinct faster than completely unconnected subpopulations. 2. There exists a gradient between stable metapopulation formation and population synchrony, with critical transitions from no metapopulation to metapopulation to synchronization, the latter frequently inducing metapopulation extinction. 3. Synchronization patterns emerge through time, resulting in synchrony groups and chimeric populations existing simultaneously. 4. There are two distinct mechanisms of synchronization: i. extinction and rescue and, ii.) stretch reversals in a modification of the classic chaotic stretching and folding.**

**Contents**

**1. Introduction.** There is a fundamental contradiction in metapopulation theory. On the one hand, general qualitative understanding assumes that local isolated populations tend to go extinct but that if they are interconnected by migration, a collection of such "subpopulations" could persist indefinitely, the very meaning of metapopulation. On the other hand, a substantial literature notes that if populations are coupled, repeated low points in population densities may become synchronized so that if, at a particular point in time, an extinction force visits one such population, it will visit all -- i.e., extinction of the entire collection of subpopulations (the metapopulation) is expected (Fox et al., 2017; Wang et al., 2015). It may seem that metapopulation theory thus must conclude that interpopulation migration is "stabilizing," but only to some level of migration, after which it is "destabilizing," the two forces apparently in opposition, but interacting in sometimes complicated ways (Abbott, 2011). A successful metapopulation thus must strike this natural balance (Griffen, and Drake, 2008). Herein we demonstrate that, at least through the lens of discrete iterative single species equations, complications may arise associated first, with the form of extinction and second, with the dynamics of synchrony.

    Extinction of local subpopulations is a core idea of metapopulation theory, whether in its original mean field form (Levins, 1969) or in the many enriching embellishments subsequently proffered (e.g., Hanski 1998; Gilpin 2012). It remains core to the very idea of a metapopulation, providing the important intuition that migration and extinction combine to determine whether a population persists or not. The evident connection with the Allee effect has also been noted (Amarasekare, 1998). Although empiricists sometimes view the Allee effect as rare in nature (e.g. Gregory et al., 2010), from a theoretical point of view local extinction of subpopulations does not occur without an Allee point (allowing an Allee point of zero is an obvious option to maintain generality). And, of course, if local extinctions of subpopulations do not occur, metapopulations do not exist.

    There is an extensive literature dealing with the extinction process. Two approaches seem evident; first, the obvious idea that stochastic forces are likely to result in extinction especially in rare populations (Lande, 1993), and second, extinction emerges from dynamic and deterministic population forces (Schreiber, 2003). Deterministic population extinction emerges in either of the classical predator prey models (i.e. Lotka/Volterra or Nicholson/Bailey), and it seems to be tacitly assumed that single population models produce extinction only when population growth rate is less then 1.0 (Gaggiotti and Hanski 2004). Yet with a combination of the Allee effect and either chaotic or intermittent populations in a discrete time framework extinction occurs with an obvious

mechanism -- if the minimum population size in a chaotic (or intermittent) attractor is less than the Allee point, the population will eventually go extinct (Schreiber, 2003; Vandermeer, 2020; 2021) with no stochastic force needed. A rich theoretical literature is relevant, acknowledging that the Allee point is actually a separatrix of the basin of attraction of the zero point, and that extinction is thus inevitable if the attractor includes reachable points below that separatrix (Grebogi et al., 1982; 1983; Vandermeer and Yodzis, 1999; Zotos et al., 2021). Here, we explore this general assumption. We note that if extinction is to arise from deterministic forces, we must assume either 1) all subpopulations are declining all the time or 2) some form of intermittent behavior is exhibited by the subpopulations such that an extinction threshold (either zero or an Allee critical point) is expected to be breached at some time.

The idea of population synchrony has likewise become conventional wisdom in ecology. From elementary considerations of classical equations (Vandermeer, 1993; 2006) to more thoughtful considerations of ecological interactions in general (Platt and Denman, 1975), ecological populations under a variety of circumstances behave like other oscillators in nature -- they form synchrony patterns (Strogatz, 2012). In practice, numerous cases of phase locking have been reported from natural populations (Benincá et al., 2009; Blasius, et. al., 1999; Blasius and Stone, 2000a; Earn et al., 1998) and a variety of theoretical formulations reinforce the basic idea ( Koelle and Vandermeer, 2005; Goldwyn and Hastings, 2008; Nobel et al., 2015; Ahn and Rubchinsky, 2020; Azizi and Kerr, 2020).

We consider a metapopulation as a collection of "propagating sinks" to use a category from the body of literature generally called source/sink populations (Pulliam, 1988; Vandermeer et al., 2010). By definition each subpopulation is doomed to local extinction, but sends out propagules before the extinction sets in. A metapopulation is thus a collection of propagating sinks. Intuitively, the Levins result notes that extinction rates must be smaller than migration rates for the whole metapopulation to persist. Each subpopulation (propagating sink) is destined to local extinction, but if migration among subpopulations is larger than extinction, the collection of subpopulations may persist in perpetuity. Yet, the coupling imposed by the inter-subpopulation migration also implies synchrony of populations, which implies eventual extinction of the whole metapopulation.

The extinction process is well-appreciated as containing a variety of complicating issues. Nevertheless, much of the literature seems to follow Darwin's simple observation that "Rarity, as geology tells us, is the precursor of extinction" (Williamson, 1989). While Darwin and his successors emphasize the extinction of entire clades, ecological dynamics are concerned with local extinctions, relevant to larger issues such as island biogeography (MacArthur and Wilson, 1967; Losos and Ricklefs, 2009), conservation (Lande, 1998), and, most importantly for this paper, metapopulations. In all applications, extinction is frequently tied in, perhaps only tacitly, with the idea of being rare (Hartley and Kunin, 2003). One gains concern in conservation, for example, when a species is thought to be rare enough to pass some lower threshold and thus be in danger of extinction. Rabinowitz's (1981) classic framework of seven types of rarity, grew from the fundamental insight that rareness happens for a variety of reasons and categorizing those reasons could aid further research into the topic. Much subsequent attention to rarity had (and still has) to do with the implicit assumption that populations headed for extinction are likely to be rare (Harnik, et al., 2012). As repeatedly noted, both empirically and theoretically, this assumption is not necessarily true (e.g., Wayne et al., 2015). Yet it is generally thought to be obvious that, at least at a local level, a rare population when subjected to some stochastic variability or environmental change, is more likely to disappear than a common one.

While empirical attention to the fact of this matter is justified and common, less attention has been paid to the nature of the extinction process itself, a potential focus that would seem similar to the insights of Rabinowitz in her seminal article. Basic population dynamics, would suggest, from both theoretical (Gottesman and Meerson, 2012) and empirical evidence (Harrison, 1991) that there are two major routes to extinction. First, in "slow extinction" a population gradually declines, as it would if living in a relatively hostile environment, but effectively hanging on through a population growth rate that is only slightly less than unity, eventually becoming extinguished as it slowly approaches zero. Such is, for example, likely the case when a population exhibits an "extinction debt," a seemingly healthy population that is slowly declining and eventually will disappear, a concept popular in conservation with concern about loss of entire species (Tilman et al., 1994). It is also a concept clearly relevant to local extinction of local subpopulations of a larger population (Hanski and Ovaskainen, 2002). Second, in "catastrophic extinction" a relatively large population suddenly collapses (e.g., a critical transition – Sheffer et al., 2016), as might be typical of a large, homogenous and dense population suddenly subjected to an epidemic disease. External events, such as asteroids or urban development, are often associated with such collapse, but extreme predatory pressure, or, especially, infectious disease is frequently implicated (Tuohy, et al., 2020).

The key distinction between these two types of extinction processes is that the first type necessarily implies rarity on the way to extinction, while the second type implies catastrophic declines from the position of a large population (e.g., a devastating epidemic is not likely with a small dispersed population). Although it is understood to be real, this second class of extinction, the catastrophic form, is less studied since it is thought to be largely unpredictable in nature. Here we are concerned with the degree to which these types of extinction processes, slow versus catastrophic, have an effect on the formation of a metapopulation as defined by a group of interconnected propagating sinks. Each of those sinks are, by definition, doomed to local extinction and the question posed is whether the type of extinction impacts the dynamics of metapopulation formation and persistence of the metapopulation.

The basic qualitative idea for both extinction forms is illustrated in figure 1, which also illustrates the basic operation of the iterative models used in this paper.

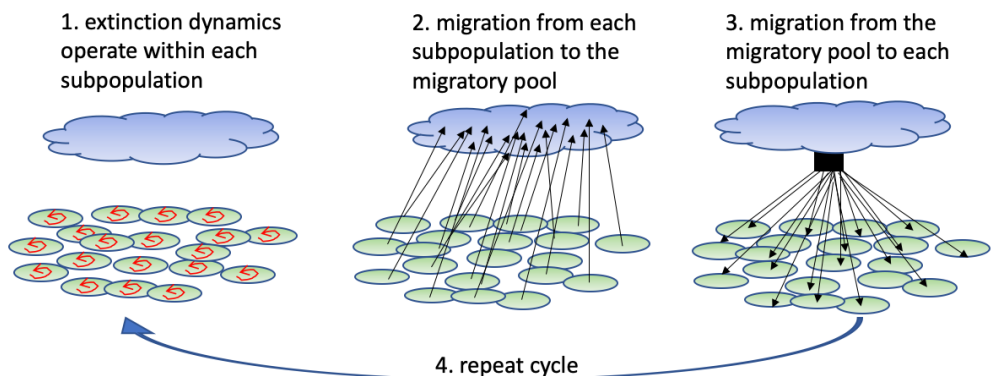

*Figure 1. Basic structure of the metapopulation framework used in this work. Small ovals and arrows indicate local dynamics within each subpopulation. Straight black arrows indicate migration (dispersion).*

In general, we are concerned with first, the extinction process and second, the synchrony patterns. We interrogate the idea that metapopulation dynamics ultimately sits between these two forces. Using two iterative map formations, the logistic/Allee map and the sawtooth map, we explore the deterministic details of the extinction process first as an underlying force and second as an emergent property of synchrony.

**2.0 The logic and dynamics of the logistic/Allee map.** Iterative maps of populations have provided basic insight into issues such as chaos, the popular logistic map being just one such example. Allowing for a critical Allee effect, we write what we refer to as the logistic-Allee map (Vandermeer and Perfecto, 2019) as,

$$X(t+1) = r(X(t) - X_{crit})[1-X(t)] \qquad 1$$

where $X_{crit}$ is a parameter that stipulates the existence of an Allee point ($X_A$), the population density above which the population must be located, in order to persist, and is given by,

$$X_A = -\frac{1-r(X_0+1)-\sqrt{1-2r(X_0+1)+r^2(X_0-1)}}{2r}.$$

There are two avenues whereby equation 1 stipulates an unviable population (i.e., the inevitability of extinction), as illustrated in figure 1. For large rates of population increase, r, (Fig 2a,b,c), a chaotic state emerges, with the boundary of the chaotic attractor and the boundary of the zero basin (the Allee point) intersecting, which means that from its peak possible density the population can suddenly crash (as illustrated in Fig 2b,c). This catastrophic extinction process, at least when modelled with equation 1, may entail a long transient period of chaos before the crash (Fig 2b). In contrast, for small rates of population increase, r, when near a subcritical saddle node bifurcation (Fig 2d), the population will drops off monotonically to zero, the reason for which is evident (Fig 2d, e).

Thus, with the logistic/Allee model we see a reflection of the two dynamic extinction possibilities discussed earlier. In what follows we refer to the first type of extinction as catastrophic drop and the second type as gradual decline.

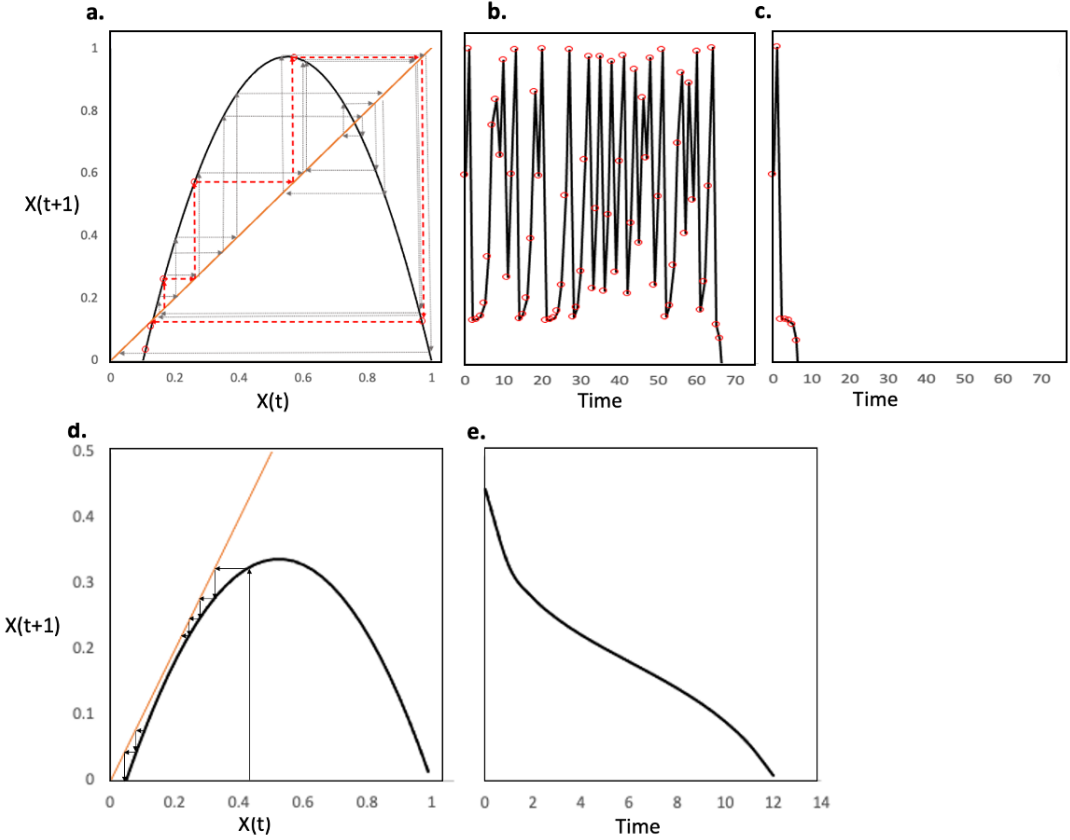

*Figure 2. Basic behavior of the logistic/Allee map. a. the chaotic situation, with parameters Xcrit = 0.1, r = 4.8. Two trajectories are shown on the same map and plotted as time series in parts b and c. b. Time series from map in a beginning at 0.578. c. Time series from map in a beginning at 0.577. d. the subcritical state with Xcrit = .05, r=1.5. e. time series from part d.*

Consider equation 1 as representing each of a collection of equations tied together by migration. That is, we presume a collection of subpopulations, each of which is subjected to the local dynamics of equation 1 with parameters set such that extinction is inevitable. Each population sends migrants out to the general region at rate "*m*", which is to say a fraction *m* of X will leave each population, creating a pool of migrating individuals. Thus, in a metapopulation consisting of N propagating sinks (local subpopulations which will go extinct in the absence of migration), we rewrite equation 1 as:

$X_i(t+1) = r(X_i(t) - X_{crit})[1 - X_i(t)] - mX_i(t) + \frac{1}{N}\sum_j mX_j(t)$  2a
$X_i(t+1) = 0$, for $X_i(t+1) \leq 0.0001$  2b
$X_i(t+1) = 1$, for $X_i(t+1) \geq 1$  2c

Note that in the present work all calculations assess the local dynamics operates first, followed by migration over the whole collection of subpopulations. The arbitrary value of 0.0001 is set as an indication of extinction.

In contrast to much of the metapopulation literature in which extinction is simply set as a fixed parameter, we here model the extinction process dynamically and thus have no explicit term for the extinction rate. Migration, on the other hand, is treated as a fixed parameter. Setting parameters to explore the implications of the two different types of extinction, either extinction via rarity or extinction via catastrophic decline, we examine the "time to extinction" of the entire metapopulation. We presume, as is central to the classical formulation, that a true metapopulation will have an infinite time to extinction, but for practical purposes any relatively long time to extinction might be also thought of as a successful metapopulation. In our formulation here, a metapopulation is considered to be extinct only when all N populations from equation 2 are extinct simultaneously.

**2.1 The emergence of a metapopulation with catastrophic extinction.** The overall dynamic diversity of this mathematical metaphor is complex and well-known. Of concern here is not its diversity of dynamical behavior, but rather its utility in studying the process of extinction in particular**.** The parameters of equation 2 can be set such that either 1) a gradual decline (Fig 2d,e) or 2) a catastrophic drop (Fig 2a, b, c) will characterize the extinction process in the unconnected populations. Setting them such that all separate subpopulations will catastrophically collapse (r = 4.8, $X_{crit}$ = 0.1), we calculate the "time to extinction" of a metapopulation with 20 subpopulations, for the full range of migration coefficients, *m*. Results are illustrated in figure 3. Note that there is a range of migration coefficient values for which there is little evidence of a metapopulation structure (from *m* = 0 to m = approximately 0.004, Fig 3a), and then, suddenly the metapopulation structure emerges. All migration coefficients between *m* = 0.004 and 0.5 include cases of time to extinction greater than 20,000 (Fig. 3a), but for *m* > 0.41, examples emerge of time to extinction far less than that, including some that become extinct more rapidly than if the twenty subpopulations had been completely unconnected (Fig 3c). We allow for the approximation that a time to extinction greater than 20,000 units can be regarded as a successful metapopulation (although this issue may be complicated, as discussed below).

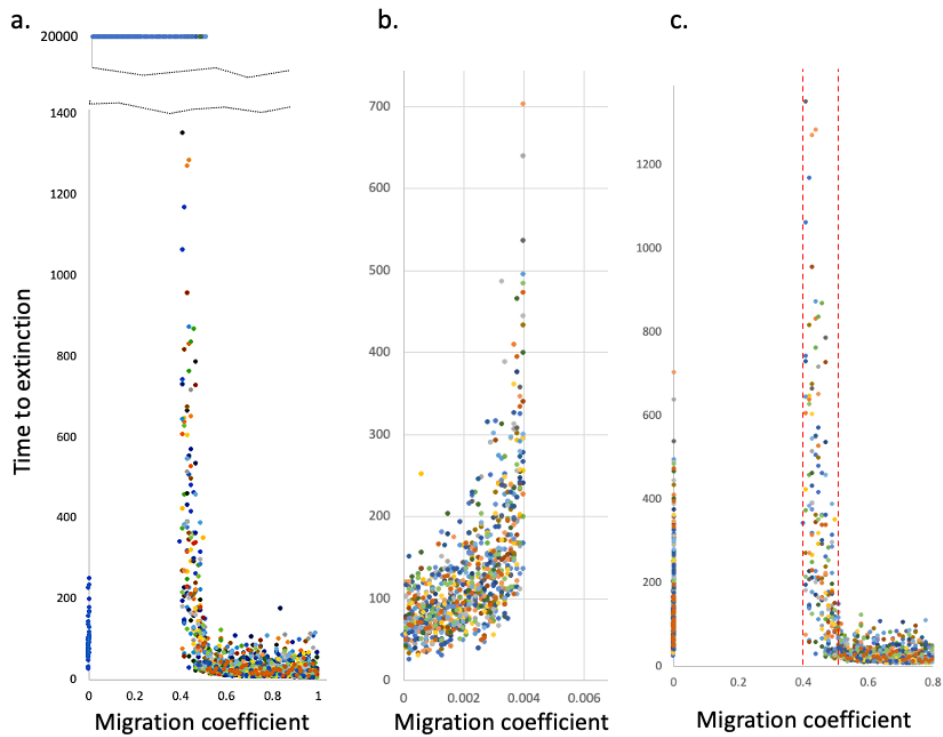

*Figure 3. Time to extinction as a function of migration coefficient for the logistic/Allee map under conditions of chaos (see Figs 1a, b, and c). Parameters are r = 4.8, and $X_{crit}$ = 0.1. a. results for complete range of migration coefficient, b. results displayed for m = 0 - 0.006.  b. c. results displayed for m = 0 – 0.8, and time to extinction from 0 to 1300, illustrating the hysteretic window (between the vertical dashed red lines).*

Consider a more fine-scale examination with the migration coefficient of *m*=0.002, in the zone where time to extinction is clearly on average greater than an unconnected group of subpopulations, but for which the time to extinction is still relatively short (Fig 3b). A typical time series (initiating all subpopulations with random numbers from a uniform distribution with range 0 – 1) is illustrated in figure 4a-c. In figure 4a, by the second iteration, one of the 20 subpopulations has descended below zero based on the logistic/Allee map, and was thus set to zero (i.e., went extinct locally). Similarly, at time 3 another of the subpopulations has gone extinct. In figure 4b, five other subpopulations go locally extinct, and in figure 4c an additional extinction occurs. Thus, by time 15 there are eight subpopulations that have gone extinct locally (descended to either zero or less than zero by the deterministic logistic/Allee map). Again, a migration pool, *P*, exists and each of those zero subpopulations will receive P/20 individuals which, on the one hand will be quite small but, on the other hand will be the same for all of the subpopulations that had descended to zero (had become locally extinct). Thus, although we describe them as having "gone extinct," the act of going extinct locally effectively forces all of those subpopulations to be in synchrony since they all receive the same fraction of the global migratory pool, and that synchrony will persist in perpetuity. The complete time series for this example is shown in figure 4d, where the time course of the "accumulation of rarity" can be seen. The time sequence for accumulating subpopulations into this near-zero synchrony group is shown in figure 4e, whence it is clear that local extinctions are rapid initially and then taper off, as reflected in figure 4e. A closer look at the pattern of synchrony-formation through local extinctions is presented in figures 4f and g. Note that by time = 80, the zero valued subpopulations occasionally do not receive a sufficient number of migrants to ascend above the critical 0.0001 level required by the model to be defined as > 0. It is this gradual accumulation of forced zeros that eventually causes all subpopulations to go extinct, signaling, by definition, the extinction of the whole population.

Thus, for this model, the formal metapopulation classification persists for almost 100 time steps. However, the distribution of subpopulation sizes in this metapopulation begins with a random distribution (by design) and rapidly approaches a situation where three subpopulations are fluctuating chaotically and 17 subpopulations persist near zero, persistent only because they are fed by migrations from the three non-zero subpopulations, all three of which persist until almost 100 time steps in obvious unconnected chaotic attractors. It is in effect a temporary situation in which the three non-zero populations are acting as sources for the 17 propagating sinks (between time 42 and 80), but is a structure that self-organizes from a model uniformly applied to a set of randomly initiated populations. We suggest referring to this structure as a pseudo-metapopulation, since it has characteristics of a metapopulation, but also of a source/sink population, where the source is clearly ephemeral (Vandermeer et al., 2010).

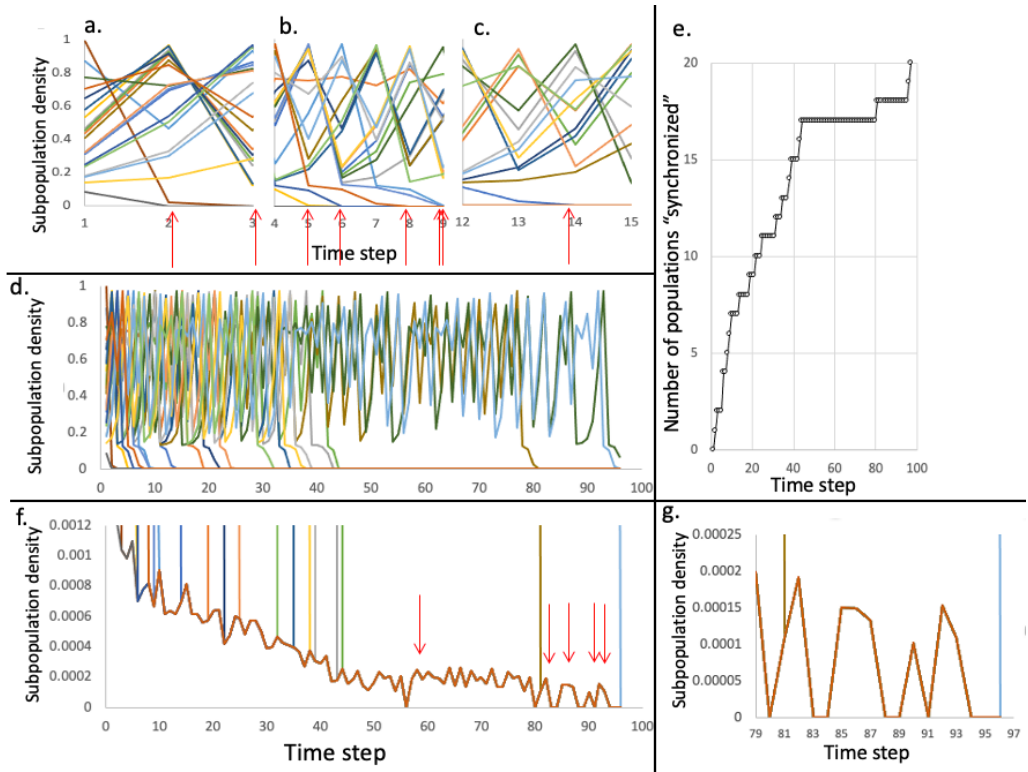

*Figure 4. Time series with migration =0 .002. a. time series for first three steps, where the one population randomly situated below the Allee point goes extinct, due to its initiation below the Allee point, and a second population goes extinct at time =3. Vertical red arrows indicate time points of subpopulation extinction. b. time series for steps 4 – 9, where five other extinctions occur, from two populations that descended to below the Allee point at time = 3, due to the underlying population dynamics stipulated by the logistic/Allee map. c. time series for steps 6 – 9, where five other extinctions occur, again due to the dynamics of the logistic/Allee map. d.) the full time series illustrating the time positions of each of the eventual 20 extinctions (synchronizations), and a time to extinction of 98. e. the cumulative number of subpopulations synchronized (initially extinguished but uniformly saved by migration) as a function of which time step the extinction occurred. f. Enhanced view of the dynamics of the rare subpopulations. Note that once the population descends below the Allee point it remains there, for these parameter values. At each iteration, those populations descend to below the critical value of 0.0001, which, according to the model (see equation 2b) it is set equal to 0. Because it (each of the subpopulations below the Allee point) receives a fraction of individuals from the migration pool, it is, by definition not extinct, although from a practical point of view it remains below the Allee point forever. g. a closer view of the later time series, showing how the in-migration from the general pool occasionally is small enough that a population near zero is not rescued from extinction (at time 83, for example, the in-migration for all the near-zero populations is not large enough to raise the population above the 0.0001 minimum, as for times 88, 91 and 94).*

At increasing migration coefficients (e.g., Fig. 3b) there is clearly a critical point at which a permanent metapopulation structure emerges ($m$ approximately 0.004). That this should occur is evident from a cursory examination of the model equations. If we allow that subpopulation subscripts are ordered according to population density (i.e., $X_1 < X_2 < \ldots < X_N$),

$$X_i(t+1) = r(X_i(t) - X_{crit})[1 - X_i(t)] + m[-X_i(t) + \overline{X(t)}] \qquad 3$$

It is evident that $\overline{X(t)} - X_i(t) < 0$ if $X_i$ is large, and $\overline{X(t)} - X_i(t) > 0$ if $X_i$ is small. Thus, the effect of migration is to decrease large projections of X, and increase small projections of X. We illustrate this effect in figure 5.

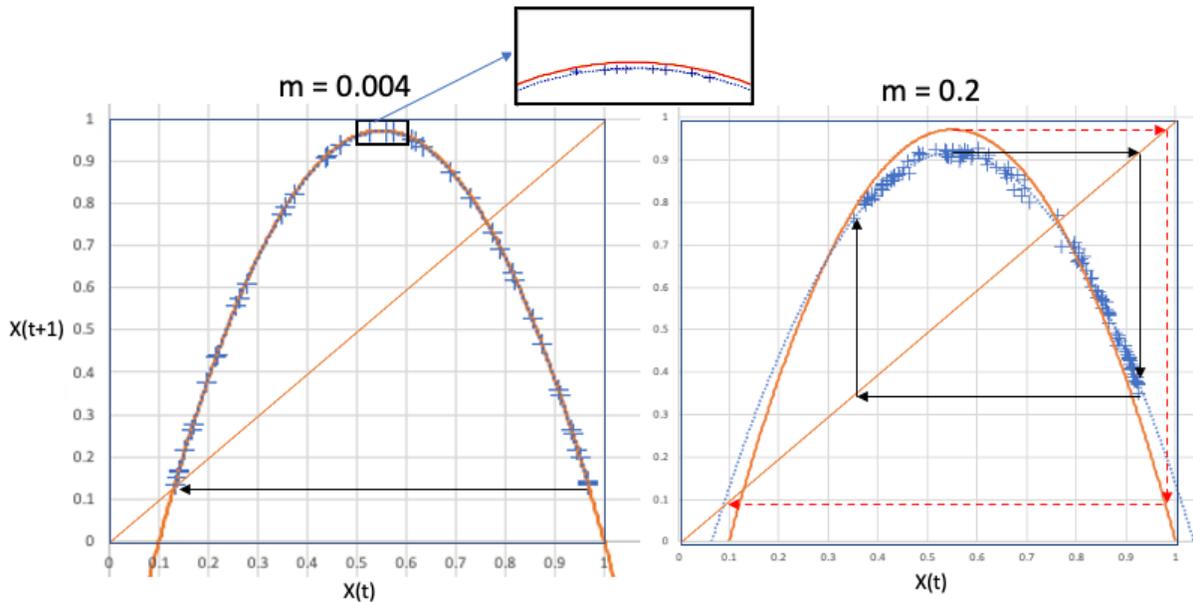

Figure 5. *Illustration of depression of high values and acceleration of low values by migration, leading to the success of the metapopulation. Red curve is the logistic/Allee kernel and plus marks are points of a single subpopulation when dynamically operating within the context of the entire 20 subpopulation metapopulation system. For m = 0.004, a very small migration rate, the actual population projections fall closely along the underlying logistic/Allee background. Yet, even here, as illustrated in the inset rectangle, the actual points are depressed from the peak of the theoretical curve. When migration is larger, as in the right hand panel (m = 0.2), the deviation of the metapopulation model (i.e., all 20 subpopulations connected) is more evident, including the obvious result that the subpopulation never reaches the Allee threshold. Dashed red arrows trace the trajectory from the peak of the theoretical curve to its intersection with the Allee point. Solid black arrows show a similar trajectory (from the peak of the empirical points) does not intersect the Allee point.*

With migration greater than $m=0.004$, a metapopulation-like structure inevitably emerges that seemingly persists indefinitely, and where the dynamics of the persistence has a particular structure. For $0.004 < m < 0.4$, time to extinction is always effectively infinite, or a dual situation emerges with one alternative a low time to extinction and the other effectively an infinite time (this alternative state occurs for $0.4 < m < 0.5$; Fig 3c). The time series that emerge within the metapopulation zone are varied. The phenomenon of synchrony is frequently (seemingly always) involved, and usually 1 – 3 synchrony groups are formed, with the important distinction that descent to below the Allee point is not the mechanism that produces synchrony. Indeed, migration operates to raise the effective recovery rate of all subpopulations to be greater than the Allee point. When a single synchrony group composed of all 20 subpopulations emerges, the entire metapopulation functions as if it were a single population and follows the "migration modified" return map. Alternatively, for some parameter values a chaotic pattern easily emerges. For example, the return map in figure 5b fits a quadratic map almost perfectly, with parameters that are close to, but clearly not the same as the original map (which remains the kernel for these trajectories). Indeed, plotting any of the two of the subpopulations against one another, a clear signal of a chaotic attractor emerges. However, extracting the quadratic fit to either of the subpopulations (an almost perfect fit), and calculating directly from the quadratic map, as if the two populations were independent (but follow the same general rule as the subpopulations in the completely connected model), it is evident that there is a regular (but either chaotic or quasiperiodic) trajectory recovered. Broadly speaking the patten is either symmetric or anti-symmetric, depending on starting values, but is constrained to a restricted basin. In figure 6 we plot both the reconstructed quadratic maps (two examples, one symmetric [red] and one anti-symmetric [black]) and the chaotic attractor actually obtained, appearing as a cloud of points surrounding the reconstructed attractor. As worthy of further study as this structure is, the important fact for the present paper is that both the reconstructed attractor and the actual attractor are clearly buffered away from the Allee point, which is the basic mechanism of metapopulation formation.

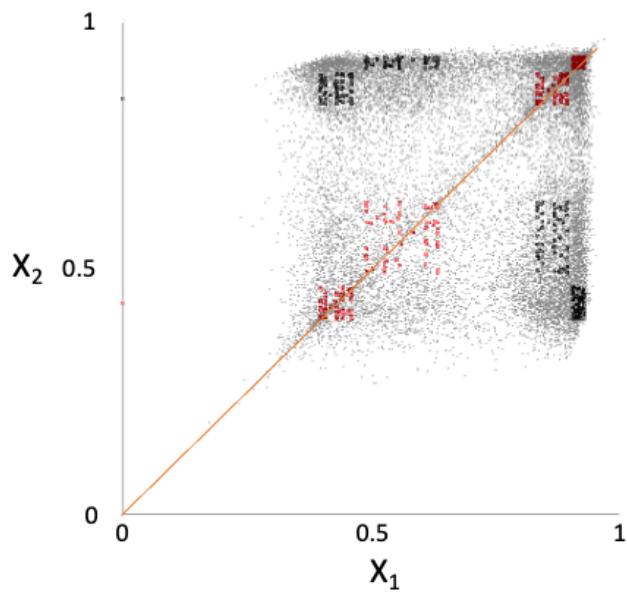

*Figure 6. With m = 0.2, a chaotic or quasiperiodic attractor emerges, envisioned with two arbitrary subpopulations (any pair has the same configuration). Solid points are trajectories from two independent runs of two populations behaving according to the reconstructed quadratic map (see quadratic fit in figure 5b), transparent points are the positions of two subpopulations in the context of all 20 subpopulations operative (any two of the 20 produce the same pattern).*

Other parameter settings within the metapopulation window, produce other complications. For example, at m = 0.28 a typical trajectory is displayed in figure 7. The first 100 time steps are displayed in figure 7a, where it is evident that around t = 70 a clear pattern begins to emerge. At this point the populations are forming into three synchrony groups (one with 6 subpopulations, two with 7 subpopulations), and the return maps for each of the three synchrony groups are distinct, but similar (Fig 7 b, c, and d). The same pattern becomes more distinctly three specific synchrony groups as time proceeds to 400 plus (not shown), and finally becomes three perfectly symmetrical population groups in a permanent eight point cycle (Fig 8).

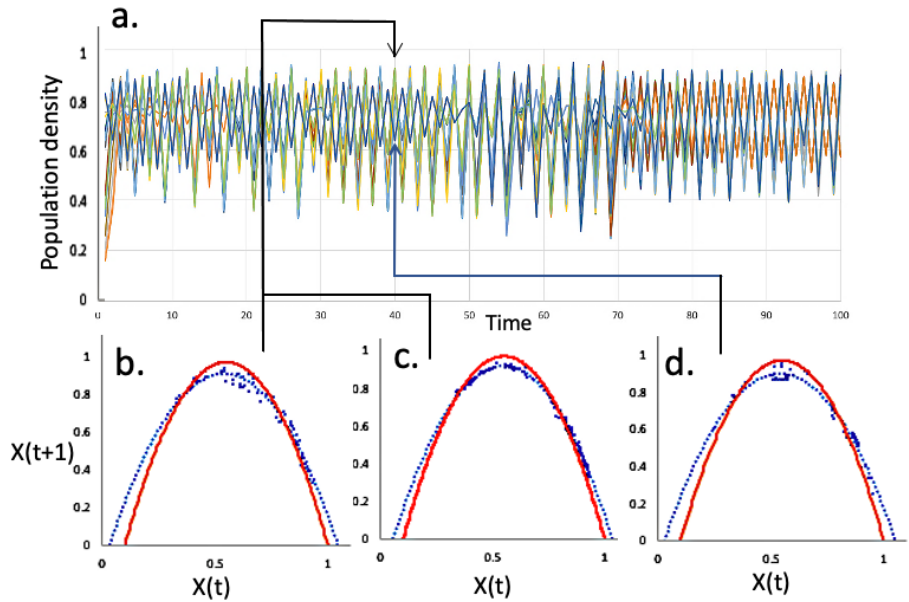

*Figure 7. Typical trajectory for m = 0.28. a. Time series for the first 100 time units. b. Return map for one of the synchrony groups composed of 7 subpopulations. Red solid line the theoretical logistic/Allee kernel, individual points are the recurrent positions of the population density of the members of the synchrony group, and the light dotted blue curve the best quadratic fit to the observed points. c. Return map for the other synchrony group composed of 7 subpopulations, points and curves as in b. d. Return map for the synchrony group composed of 6 subpopulations.*

In figure 8 we display the final stage of the trajectory displayed in figure 7, showing the final state on a return map, for each of the permanent synchrony groups.

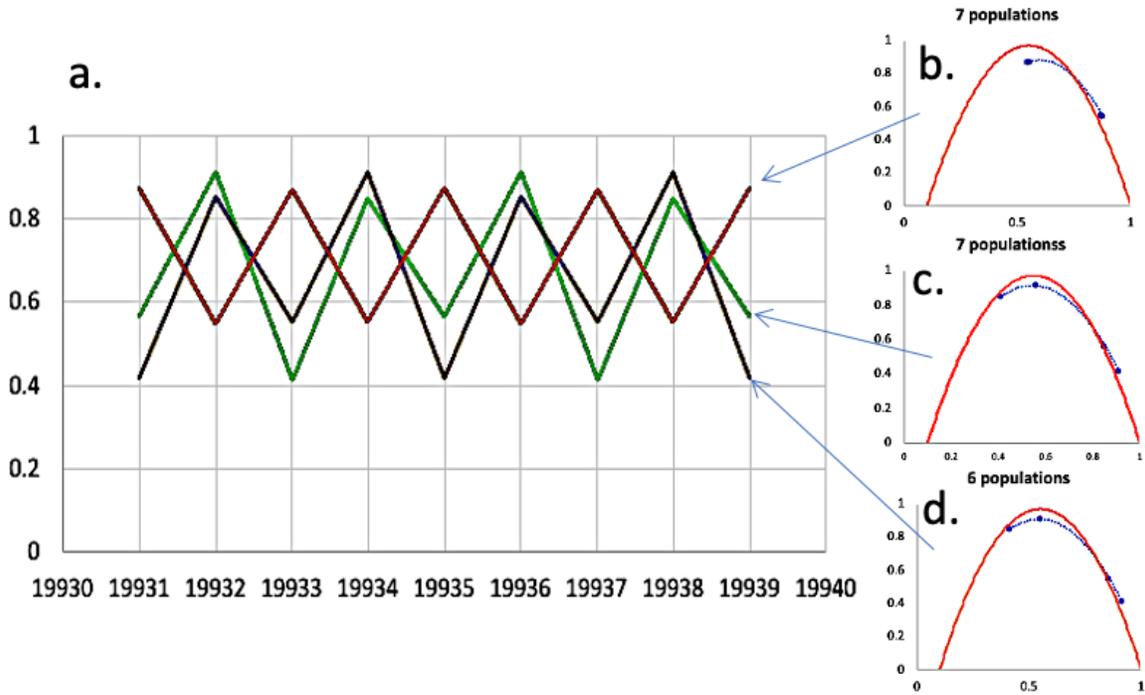

*Figure 8. Late time series (time = 19,930 to time = 19,939) for the time series shown in figure 7. Note that the synchrony groups are now completely formed and constant through time. a. the time series with all 20 subpopulations (each within a synchrony group appears to be a single population). b. return map of the two-cycle synchrony group composed of 7 subpopulations, qualitatively in reverse synchrony with the other two synchrony groups. c. One of the 7 subpopulation synchrony groups in 3-cycle synchrony with the other 7 subpopulation synchrony group. d. the other 7 subpopulation in 3-cycle synchrony.*

At the same migration rate ($m = 0.28$) but from different initiation points, the chaotic attractor with two synchrony groups (one with 12 subpopulations the other with 8 subpopulations) is displayed in figure 9. Note that the two synchrony groups are phase reversed in a relatively fixed fashion and produce a typical chaos-like structure on a return map (fig 9b). Importantly, the edge of the chaotic attractor is positioned well above the Allee point, meaning that the collection of subpopulations is clearly a stable metapopulation.

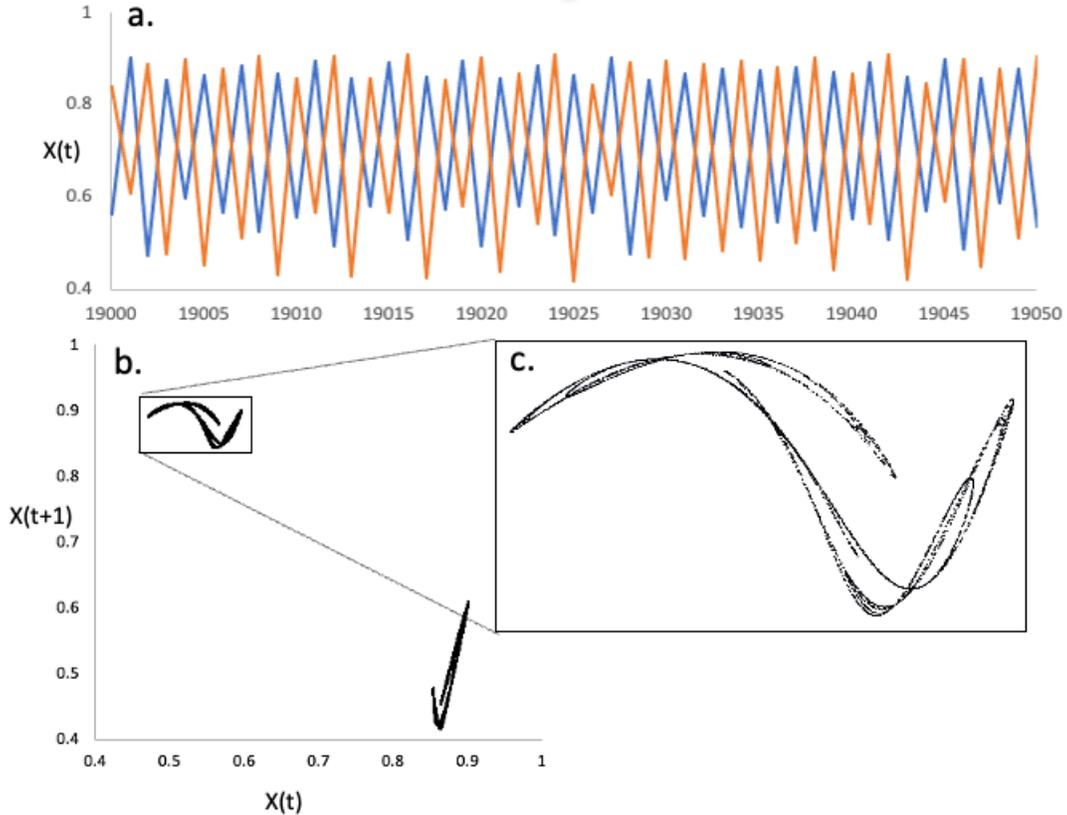

*Figure 9. At m = 0.28 from an initiation point distinct from the the simulations in figure 8 we see a fixed chaotic attractor, with its edge far removed from the Allee point. a. time series after 19,000 iterations, illustrating the two synchrony groups are, qualitatively speaking, phase reversed. b. Return map of the time series in a, showing the typical phase reversed chaos (Vandermeer and Kaufman, 1998). c. Microscopic view of the half of the attractor pictured in b.*

With these three exemplary cases (Figs 6 – 9) it is evident that a range of possible dynamic behaviors is contained in this model. Our purpose is not to explore this diversity, but rather to note that in all cases, the dynamics is contained to be above the Allee point and that frequently massive synchronization is involved. Thus, the idea that synchronization is, in and of itself, a force that cancels the metapopulation, is incorrect. It is through synchronization that the complicated structures (as illustrated in Figs 5 – 9) emerge. Furthermore, it is possible to derive a qualitative generalization for maintenance of the metapopulation through synchronization with this model, as illustrated in figure 10. In all observed cases, each subpopulation, unless zero or some constant value, follows a quadratic pattern on a return map. But that empirical pattern, as would be expected from the above discussion (see especially equation 3), is systematically distorted through the joint migration effects of all subpopulations involved. Intuitively, the cause of the well-known sensitive dependence on initial conditions is the sections of the map for which

$$\frac{dX(t+1)}{dX(t)} > |1.0| \qquad\qquad 4$$

where the standard notion that the interval between two trajectories will expand (the "stretching" of a classic chaotic attractor) is clearly seen, as the expected "stretching" of a simple chaotic attractor, as in the potent metaphor of the Smale horseshoe (Smale and Shub, 2007). However, when equation 4 is not satisfied, two nearby trajectories will contract (the opposite of the stretching of a classic chaotic attractor), a "reverse stretching" (one could also say contracting or compressing). It is evident that this is a mechanism of synchronization and is quite distinct from resetting subpopulations at zero of previous cases. If the stretching part of the "stretching and folding" mechanism (recall Smale's horseshoe) is reversed, as will be the case when the inequality of equation 4 is reversed, it makes sense to refer to this synchrony mechanism as "stretch reversal." The synchrony mechanism referred to above as "extinction symmetry" is quite distinct from this idea. In the zone of metapopulation persistence (Fig 3, between .004 and .4), it is commonly (perhaps always) the case that each subpopulation fits a quadratic function quite well, clearly a reflection of the fact that the kernel of the model is the logistic/Allee quadratic function. However, the fit is inevitably distorted, seemingly in the same direction, whatever the migration coefficient. This distortion is illustrated in figure 10. The fitted curve is platykurtic (with respect to the logistic/Allee kernel) and shifted to the left. Thus, a greater proportion of the X space is involved in stretch reversal than in the original kernel, leading to massive synchronization.

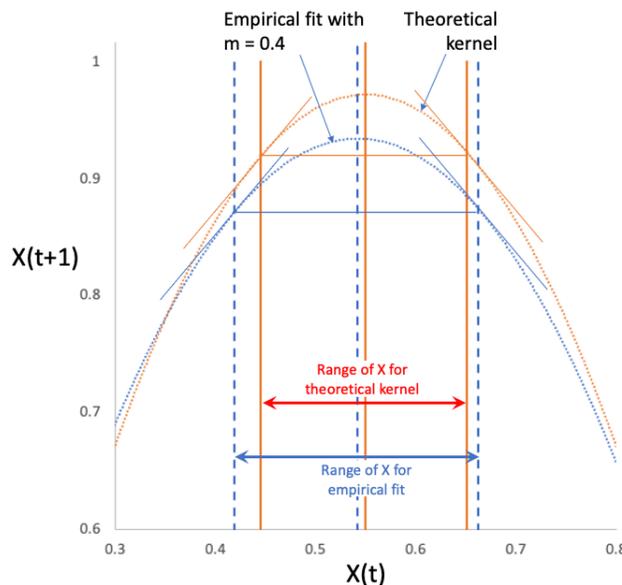

*Figure 10. Illustration of the change in relative amount of projection space leading to stretch reversal ("range of X") for both theoretical kernel and specific empirical fit. Bold horizontal lines indicate the critical values, above which stretch reversal occurs.*

**2.2 The subcritical state of the logistic/Allee map.** The other major form of extinction that is evident in the logistic/Allee map is extinction via rarity, captured by subcritical dynamics of the model (Fig. 2d, e). Applying equation 2 in the subcritical parameter region produces the results in figure 8. What may seem surprising is that the metapopulation framework does not produce a metapopulation -- time to extinction for all migration coefficients is more rapid than with unconnected subpopulations. As migration is added to independent subpopulations, even a very small migration coefficient decreases the time to extinction, and with further increases in migration, that time declines yet further. Thus, when the underlying dynamics of the extinction process is fueled by rarity (the subcritical case), metapopulation construction with migration is seemingly not possible, in contrast to subpopulations with catastrophic extinction dynamics. Upon reflection, this conclusion is perhaps obvious. If each subpopulation is set to be monotonically declining without exception, overall migrants in the system will also be continually decreasing. The emergent property is, unsurprisingly, the entire collection of subpopulations goes extinct faster with migration than without, the precise opposite of the basic metapopulation construct.

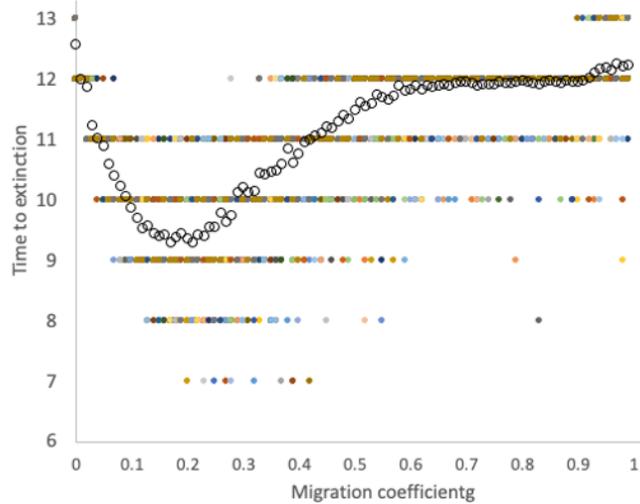

*Figure 11. Time to extinction versus migration in the case of subpopulation extinction emerging from rarity in the population dynamics, emerging from the subcritical parameter space of the logistic/Allee model (see Fig. 2d, e). Parameters are r= 1.8 and $X_{crit}$ = 0.08.*

**3 The ecological significance of the sawtooth map.** The logistic map and its many relatives (e.g., the Ricker map) have provided much insight in population biology, as noted. It carries with it two fundamental assumptions about biological populations, first, that they grow (at some initial rate, frequently referred to as the intrinsic rate, which is frequently assumed to be a constant when populations are small), and second, that if they reach a certain high density they decline. A certain elegance is contained in the model in that the switch from an increasing population to a declining one is instantiated with a smooth function. That is, a simple one-dimensional iterative process is "smooth" in its definition (a simple quadratic or similar function), yet captures what is essentially a non-continuous switch from an increasing to a decreasing population. That result accords with a common sense notion of population dynamics – populations grow until they become overpopulated and then they decline. Yet that convenient elegance carries with it a result that is not necessarily all that realistic, at least not as a generalization.

The decline of the population is of two types as addressed above. First, sometimes a population declines catastrophically, as when struck by a disease epidemic, for example. As illustrated in figure 2a -c, one parameter set generates a population increase followed by what is frequently a catastrophic drop. A completely different parameter set generates a smooth population decline to zero as illustrated in figure 2d,e. There is no parameter combination that alone can accommodate both rapid increase and smooth decline. In other words, the model cannot accommodate a pattern of a single population operating according to constant parametric rules gradually increasing to a point and then transitioning to a gradual descent. Yet that is a pattern frequently suggested in nature. On the one hand, when the intrinsic growth rate is large, population values exceeding some threshold will "crash" and, the larger the population preceding the crash, the smaller the population resulting from that crash. The internal dynamics imparts the rule that the population resulting from a decline will be small if the state it comes from is large. In other words, a population this year is negatively associated with its previous year's density (Fig 2a – c). On the other hand, those cases in which declining populations approach zero as if approaching any other equilibrium point, smoothly declining rather than crashing suddenly from a previously large population, must be accommodated with a distinct parameter setting (Fig. 2d,e).

Other models exist in which population trajectories can both gradually increase and gradually decrease with the same parameter settings. For example, the classic Bernoulli map (Nee, 2018),

X(t+1) = 2x             *Mod 1*                                              5

a graph and exemplary trajectories of which are indicated in figure 12a, is one such example. A useful modification of the Bernoulli map is the Pomeau/Mannville map (Klages, 2013; Nee, 2018; Vandermeer, 2019; 2021),

$X_i(t+1) = rX_i(t) + aX_i(t)^b$         *Mod 1*                                  6

a graph and exemplary trajectories of which are indicated in figure 12b, where *X* is meant to symbolize population density or population biomass, *r* is the unencumbered growth of the population, *i* is the *i*th subpopulation in the metapopulation, *a* and *b* are arbitrary parameters which may be given approximate biological meaning (Vandermeer, 2021). The essential feature, from a biological perspective, is that the decline of the population toward extinction is not constrained to be "catastrophic" (emanating from an extreme value of the population density), but to most closely simulate the extinction via rarity, the smooth approach to zero, as discussed earlier.

A model with similar behavior is a simple linear "sawtooth" map, used in diverse applications (e.g., Mondragon et al., 2000; Vignoles, 1993). In figure 12d–h we illustrate the transformation of a two-stage linearization of the rising half of the logistic/Allee map to a sawtooth map that, in the end, resembles a linearized version of the Pomeau/Mannville map (Fig 12b), or a distorted version of the Bernoulli map (Fig 12a).

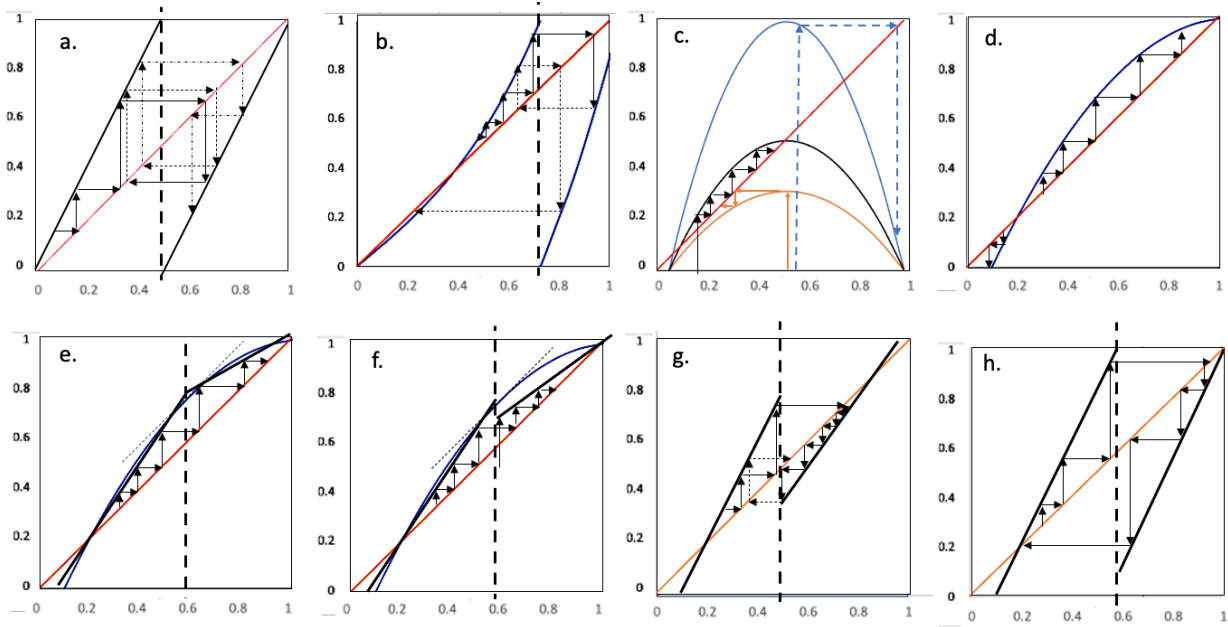

*Figure 12. Construction of the sawtooth map. a. the standard Bernoulli bit shift map, with two linear segments (equation 5). b. the standard Pomeau-Mannville map (equation 6) with r=0.85, a = 1, z = 3. c. the logistic-Allee map (X(t+1) = r(X(t) – X$_{crit}$)(K – X(t))/K) with three values for the growth parameter r = 4.4, 2.3 and 1.4, from top to bottom graphs, Xcrit = 0.05 and K = 1, for all three. It is the top function (in blue) that gives the catastrophic drop in population density and the bottom one (in gold) that gives the smooth transition to extinction. Three sets of trajectories illustrate the three distinct dynamic structure possible from this model. d. the logistic-Allee map with X$_{crit}$ = 0.1, K = 2, and r = 1.1, the shift in K is to represent the equation in the same framework as the subsequent formulations. e. piecewise linearization of the logistic-Allee map, with the break point at the inflection, where dX(t+1)/dX(t) =1 defines the inflection. Note that the qualitative behavior of the piecewise linear map is very similar to the original logistic-Allee map (in part d). f. Disconnecting the upper leg of the linear map from the lower leg. Again the qualitative behavior of the system is similar but notably different for the latter part of the increase toward X = 1. g. Disconnected linear piecewise map where the upper leg switches the population to a declining situation. h. Sawtooth linear map, which captures the qualitative behavior of the Pomeau-Mannville map of part a.*

The piece-wise Pomeau-Manville map, or the sawtooth map are alternatives that produce both chaotic-like behavior and a more sensible declining population than the logistic-Allee map. Consider the following simple sawtooth map as a model of a population in discrete time:

$$X_{t+1} = \alpha + \frac{1-x_a}{x_0 - x_a} X_t \quad \text{for all } X_t \leq x_0 \quad \quad \quad 7a$$

$$X_{t+1} = \beta + \frac{1-x_r}{1-x_0} X_t \quad \text{for all } X_t > x_0 \quad \quad \quad 7b$$

Where,

$$\alpha = \frac{x_a(x_0 - 1)}{x_0 - x_a} \qquad 7c$$

$$\beta = \left(1 - \frac{1 - x_r}{1 - x_0}\right), \qquad 7d$$

and, furthermore,

$$X_{t+1} = 1 \quad \text{for all } X_t \geq 1 \qquad 7e$$
$$X_{t+1} = 0 \quad \text{for all } X_t \leq 0 \qquad 7f$$

an exemplary graph of which is illustrated in figure 13, with the parameters fixed on the axes. The biological meaning of each of the three parameters is evident from the graph. Unlike the logistic/Allee map which has two parameters (r and $X_{crit}$), the sawtooth map has three, $x_0$ which is the critical point at which the population switches from increasing to decreasing (and vis versa), $x_a$ which is the Allee point, the population density below which the population will inevitably become extinct, and $x_r$, which is the recovery rate, the lowest point that a declining population can possibly attain. It is evident that for a sufficiently large slope of the exponentially increasing part of the function, chaos may ensue. For a random set of initial values of $X$, the time to extinction, $T_{ext}$, should be inversely proportional to $(x_a - x_r)$.

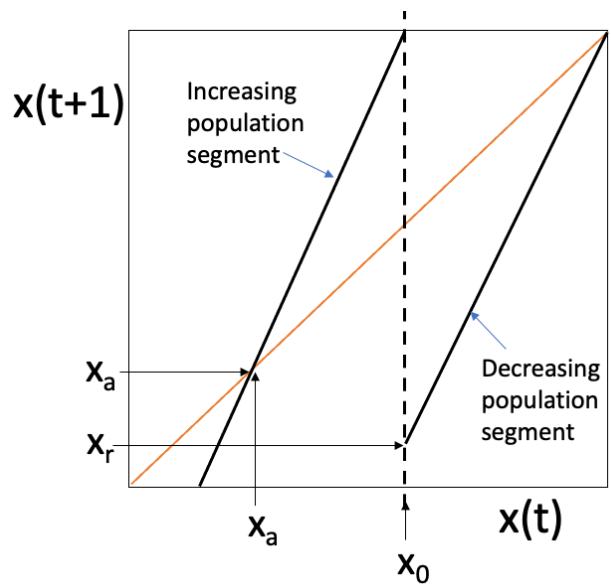

*Figure 13. Graph of equation 7, illustrating the meaning of all the parameters. Note that the parameter $x_r$ (critical recovery point) must be less than $x_a$ (critical Allee point) if the population is allowed to go extinct. Indeed, if the trajectories span the range of 1.0 to $x_a$, the population will inevitably go extinct. Not all trajectories need be chaotic since the angles of the two linear segments are not*

> *necessarily equivalent, leading to the likelihood of long periodic trajectories, similar to that found in the logistic/Allee map.*

In a metapopulation with N subpopulations, we assume that all subpopulations are connected by a migration coefficient equal to *m*, similar to our treatment of the logistic/Allee map. Thus the governing equations become, for the *i*th subpopulation,

$$X_i(t+1) = \alpha + \frac{1-x_a}{x_0 - x_a} X_i(t) - mX_i(t) + \frac{1}{N}\sum_j mX_j(t) \quad \text{for all } X_t \leq x_0 \qquad 8a$$

$$X_i(t+1) = \beta + \frac{1-x_r}{1-x_0} X_i(t) - mX_i(t) + \frac{1}{N}\sum_j mX_j(t) \quad \text{for all } X_t \geq x_0. \qquad 8b$$

retaining the other stipulations of equation set 6.

**3.1 Local extinction and synchrony patterns of the sawtooth map.** Consider N subpopulations each of which is governed by equations 8 with *m* = 0. All N subpopulations will go extinct within $T_{ext}$ time units (of course $T_{ext}$ may be very large, the meaning of a metapopulation). Connecting the populations (i.e., allowing *m* > 0) we find the time to extinction changes dramatically as migration rate increases for low critical switch points ($X_0$) (Fig. 14). For a very small critical switchpoint ($x_0 = 0.14$, Fig. 14) the unconnected subpopulations (*m*=0; Fig 14) on average has a longer time to extinction than the connected populations (*m* > 0), contrary to the underlying assumptions of metapopulation theory. Much like the case of the subcritical logistic/Allee model (Fig 14), this case of very low critical switch point ($x_0$ small) has the qualitative characteristic that the declining mode of the population ($X(t) > x_0$) dominates the domain (Fig 13), which is to say that most of the time the subpopulations themselves are in declining mode, much as in the case of the subcritical logistic/Allee case (Fig 10). And, as in that previous example, we see that the mere connecting of subpopulations through migration does not generally create a successful metapopulation. The time to extinction for *m* = 0 ranges from 36 – 245 (mean of 69 time units), whence it is clear that most of the simulations for $x_0 = 0.12$ reach metapopulation extinction much faster than would a collection of unconnected subpopulations. And the cause of this perhaps surprising result is the underlying prevalence of declining $X_i(t)$, that is, a critical switch point near to the value of $x_r$.

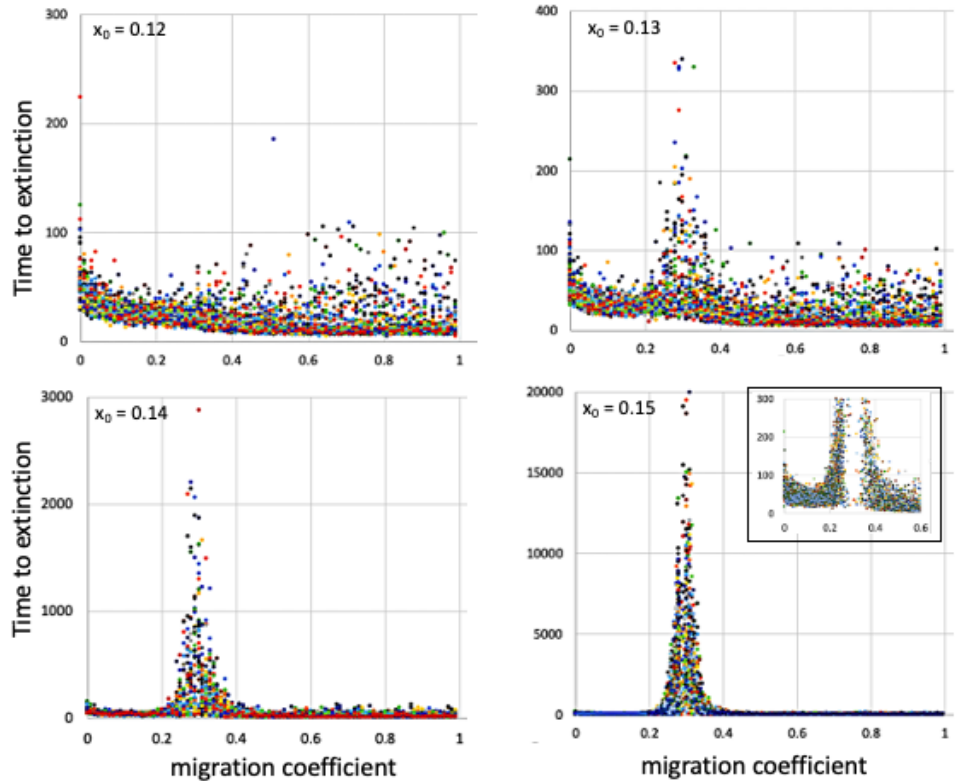

*Figure 14. Time to extinction as a function of migration coefficient for four examples of low critical switch points ($x_0$).*

As the critical switch point increases (Fig 14), there is the emergence of a zone of migration coefficient (between approximately $m=0.2$ and $m=0.4$) in which the time to extinction rapidly increases, indicating a zone of metapopulation success, at least as measured by the time to extinction (longer time to extinction indicates a more metapopulation-like behavior). Thus, as a window of metapopulation success opens up in this zone, we see the emergence of what might be called a window of metapopulation success, especially evident in figure 14 at $x_0 = 0.15$, where a critical transition to metapopulation permanence as well as a critical transition from metapopulation permanence to failure, can be seen bordering a zone of metapopulation permanence. This is evident in the inset panel in figure 14, for the $x_0 = 0.15$ case.

The initial decline in time to extinction at low $x_0$ and low m, is clearly in opposition to the basic idea of a metapopulation, as is the case of subpopulations following the rule of the subcritical logistic/Allee map (Figs 2d, 3, and 11). This pattern stems from the dual effect of the migration parameter. First, connecting populations has an effect of synchronizing them below the Allee point. By definition if left unconnected, any population below the Allee point will rapidly descend to zero. However, with a small migration coefficient, these near-zero subpopulations are rescued by migrations from the abundant subpopulations. Yet, because the model places all subpopulations below zero at exactly zero, all of those subpopulations will receive precisely the same number of migrants from the migration pool. That is, there will be one large synchrony group always below the Allee point, and that synchrony group will accumulate more subpopulations as chaotic trajectories continue "feeding" the densities below zero. As the number of subpopulations below the Allee point increases, the number of abundant subpopulations decreases so that the migration pool feeds less and less into the near-zero synchrony group. The pattern is illustrated for a particular case (with 100 subpopulations) in figure 15. As is evident (Fig 15c), the average population density of those subpopulations in the near-zero synchronization group declines as more non-zero density subpopulations descend below the Allee point, eventually reaching a point where the single non-zero population also descends below the Allee point and the metapopulation becomes extinct.

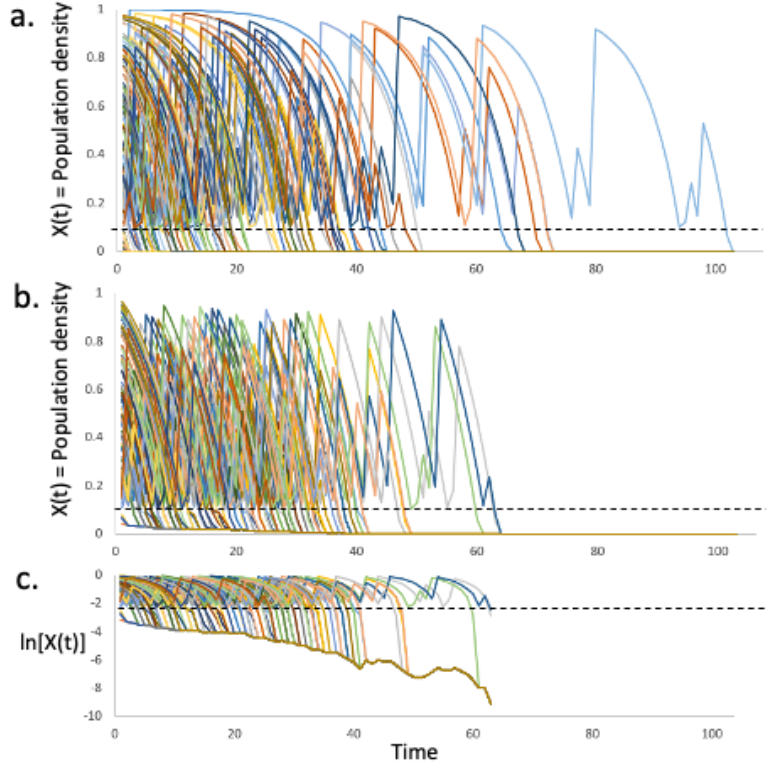

*Figure 15. Illustration of the accumulation of populations in the near zero synchrony groups. a. 100 populations unconnected. b. 100 populations connected with a migration coefficient of 0.05. c. log plot of populations over time, illustrating the sink-like nature of the near-zero synchrony group.*

As migration increases further, a distinct process comes to dominate the dynamics. As discussed above, we modify equation 7 for the i$^{th}$ species,

$$X_i(t) = \alpha + \frac{1-x_a}{x_0-x_a} X_i(t) + m\left[-X_i(t) + \overline{X(t)}\right], \text{ for all } X_i(t) < x_0 \quad \quad 8a$$

$$X_i(t) = \beta + \frac{1-x_r}{1-x_0} X_i(t) + m\left[-X_i(t) + \overline{\overline{X(t)}}\right], \text{ for all } X_i(t) \geq x_0 \quad \quad 8b$$

Note that $\overline{X(t)} < X_i(t)$ for large $X_i(t)$ and $\overline{X(t)} > X_i(t)$ for small $X_i(t)$, and differentiating 8,

$$\frac{dX(t+1,i)}{dX(t,i)} = \frac{1-x_a}{x_0-x_a} - m, \text{ for } X(t,i) << x_0 \quad \quad 9a$$

and

$$\frac{dX(t+1,i)}{dX(t,i)} = \frac{1-x_r}{1-x_0} - m, \text{ for } X(t,i) >> x_0. \quad \quad 9b$$

Thus the observed slope of the output from the model ($\frac{dX(t+1,i)}{dX(t,i)}$) with migration will be greater than the slope of the sawtooth kernel ($\frac{1-x_a}{x_0-x_a}$ or $\frac{1-x_r}{1-x_0}$). From equation 9b we see that the declining population segment of the model will fix the intercept at $x_0$ at a point larger than $x_r$, and eventually larger than $x_a$, which means that the possibility of extinction of the whole collection of subpopulations (the metapopulation) disappears. This is the origin of what we call the metapopulation window. Also, following the qualitative analysis presented in figure 10, the empirical slopes of the model (with migration) will eventually be less than 1.0, which induces the stretch reversal behavior as discussed above. In figure 16 we illustrate the gradual lowering of the derivative of the empirical points, wherein it is evident that the behavior expected from the qualitative theory of equations 8 and 9 are realized. Note that as the system moves from $m = 0.3$ to $m = 0.2$, a basin boundary collision occurs, in which the separatrix of the zero equilibrium point (the Allee point) is intersected with the boundary of the chaotic attractor. This phenomenon is effectively the source of a critical transition from metapopulation failure to metapopulation success. Compare figure 16 with figure 5, illustrating the similar biological phenomena of larger populations producing populations lower than expected and smaller populations producing more than expected. Also, the slope of the fit to the empirical points corresponds to the rule of stretching (when >1) or stretch reversal (when <1), as in the case of the logistic/Allee map (Fig. 10).

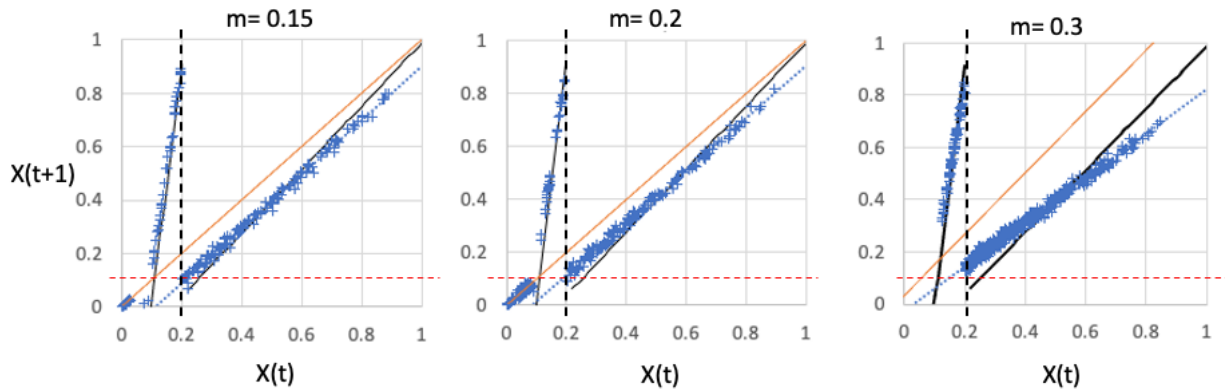

*Figure 16. Return map of model output (blue plus signs) compared to theoretical sawtooth kernel. Dotted blue line is regression of the descending set of recursion points. Horizontal dashed red line indicates the Allee point. Vertical dashed black line indicates position of $x_0 = 0.2$.*

**3.2 Metapopulation collapse at low migration.** A close examination of the patterns in figure 14 reveals an unexpected pattern. The rate of extinction (rate at which individual subpopulations go extinct) declines for low migration rates, which, contrary to normal expectations of metapopulation theory, means that migration does not cancel the expected extinction of the collection of subpopulations. Rather, the effective "below Allee sink" drags subpopulations to simultaneously transcend the Allee point periodically and thus synchronizes them (as discussed above). When two of the 20 subpopulations are synchronized, that effectively leads to the overall metapopulation containing only 19 effective subpopulations. When 15 (say) are synchronized, that effectively leads to the overall metapopulation containing only 5 effective subpopulations. The result is that the expectation of increased time to extinction with migration not only fails to materialize with migration, but there is a generally more rapid approach to metapopulation extinction as migration is increased (Fig. 17, left three time series). This result is not surprising since in this case, with $x_0$ set at 0.2, the parameter space is such that a slowly declining population defines the dynamics for 80% of the input variable in the iterative map (from X(t) = 0.2 to 1.0). As discussed above for the logistic/Allee map, when subpopulations are always in decline, the survival of the subpopulations as a metapopulation is impossible.

As migration increases yet further ($m > 0.05$), the boundaries of the chaotic attractors are gradually elevated (as in Fig. 16), such that the intersection of the boundary with the Allee point (the classical basin/boundary collision – Vandermeer and Yodzis, 1999) becomes less and less extensive, such that it takes longer for the process of accumulation of near-zero synchrony groups to begin. This begins the pattern of increasing time to extinction, as is evident in the progression of m in figure 14, and by the vertical dashed red lines in the three time series on the right hand side of figure 17.

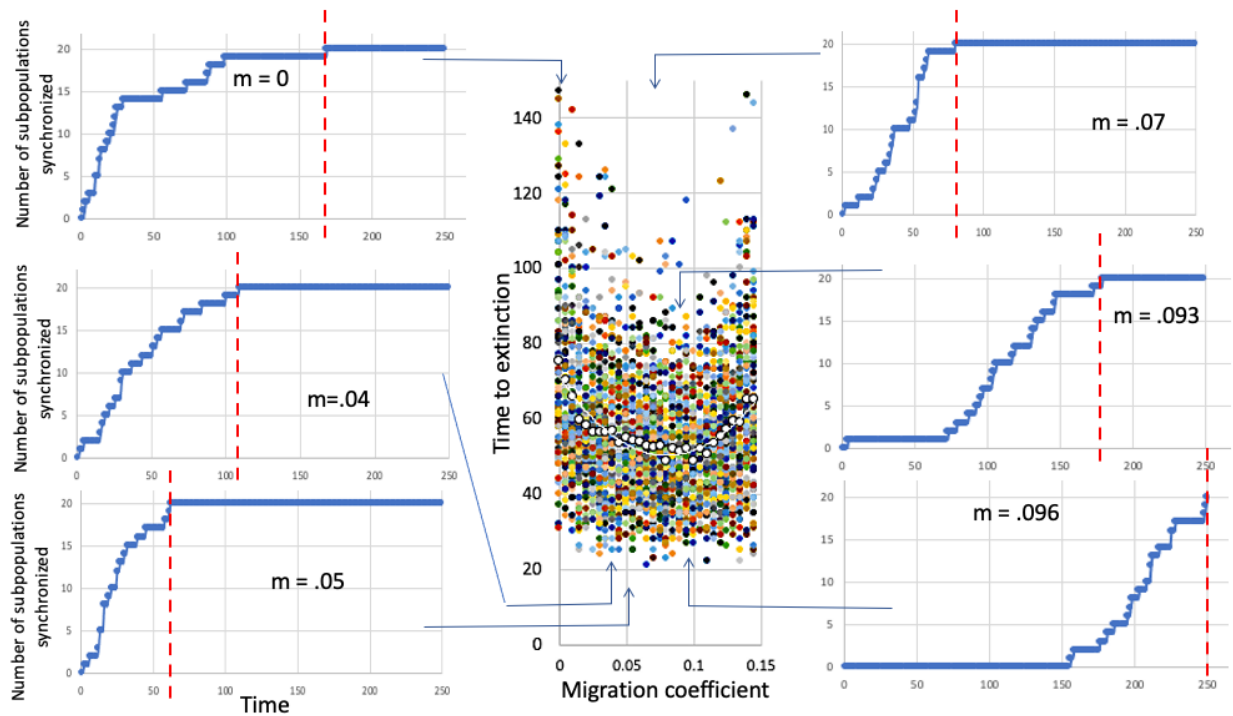

*Figure 17. Illustration of the reduced metapopulation construction at low migration levels for $x_0$ = 0.2. The central panel is a scattergram of the time to extinction as a function of the migration coefficient, the white points representing the mean of 100 separate runs and the white curve the best fit quadratic function. The three time series on the left illustrate the rate of accumulation of near-zero synchrony groups for three migration coefficient values. Note how the time to extinction (red dashed vertical line) declines with increasing migration, a function of the more rapid accumulation of near-zero synchrony groups. The three time series on the right reflect the same process as those on the left, except the effective total metapopulation dynamic is distorted as illustrated in figure 16, such that the lower boundary of each of the chaotic attractors (for each subpopulation) intersects the Allee point less frequently, meaning that there is a lag in the initiation of accumulating near-zero synchrony groups.*

The pattern in figure 17 is perhaps evident from equation set 8, but is also explicable in more qualitative terms. Consider, for example, a metapopulation of five subpopulations bound to the region 0 -1, each one of which is in a chaotic trajectory, the edge of which intersects the boundary of the basin containing zero. That is, all populations will eventually descend below the Allee point, and rapidly become extinct. Suppose the initial conditions are $X_i$ = 0.01, 0.2, 0.4, 0.5 and 0.7, and that the Allee point is 0.201. So, if the populations are completely unconnected (migration = 0), the first population will descend rapidly to zero (since it is well below the Allee point), the second through fifth, soon after, simply through the process of a chaotic attractor with a boundary that intersects the Allee point. However, now suppose that there is a small migration term, say $m$ = 0.1. The first population will descend to effectively zero, but will receive migrants from the other four populations [0.1(0.01+0.2+0.4+0.5 +0.7) ]= .141/5 = .0282, which is well below the Allee point, but the population was rescued from extinction by that small migration term. That is, the population is not zero (as would be if the basic model were followed), but rather the very small 0.0282. The next cycle, suppose the new population numbers are 0.0282, 0.0282 (the second population had been just under the Allee point, so it now it too is near extinction), 0.2, 0.7, and 0.4. This time the migration pool is 0.1(0.0282 + 0.0282 + 0.2 +0.7 + 0.4) = 0.13564, and each of the populations receives a fraction of that pool, 0.13564/5 = 0.027128. Since the two very rare populations likely descended to very near zero, their new population densities will both be 0.027128. But note, all five subpopulations continue to avoid extinction! Had there been no migration, the first two populations would now be extinct. Each time one of the subpopulations joins the collection of subpopulations that are below the Allee point, it effectively becomes completely synchronous with all the other subpopulations below the Allee point. This process continues until all the subpopulations "join" the below-Allee-point subpopulations, at which point there are no migrants to save any of them and they all go extinct. While this successive joining of the below Allee populations is occurring, the boost that all populations get from the migration pool generally, becomes smaller and smaller as they decline in unison. Thus, each of the remaining subpopulations is more likely to descend below the Allee point at every cycle. Consequently, contrary to the situation of zero migration, subpopulations will on average remain non-extinct longer, but the population as a whole will go extinct faster. Clearly this situation may arise from particular parameter settings (such as those used to generate figure 17) and we do not suggest that the situation is universal. However, it seems likely that in the gradual increase of the migration parameter in general, before a true metapopulation stability is obtained, this intermediate stage of "lowered subpopulation extinctions but higher metapopulation extinction" is likely to happen. According to the sawtooth model, this curious result emerges when population decline occupies much of the potential phase space, which can occur in either the sawtooth situation or the logistic-Allee gradual decline situation.

**3.3 The dual nature of synchrony and the "metapopulation window".** Extending the simulations of figure 14 to explore the patterns generated by distinct values of $x_0$, the general picture of a critical transition to metapopulation permanence as well as a critical transition from metapopulation permanence to failure, can be seen bordering a zone of metapopulation permanence. This pattern of a metapopulation window occurs, but with different structure for all parameter settings of the sawtooth map (Fig 18).

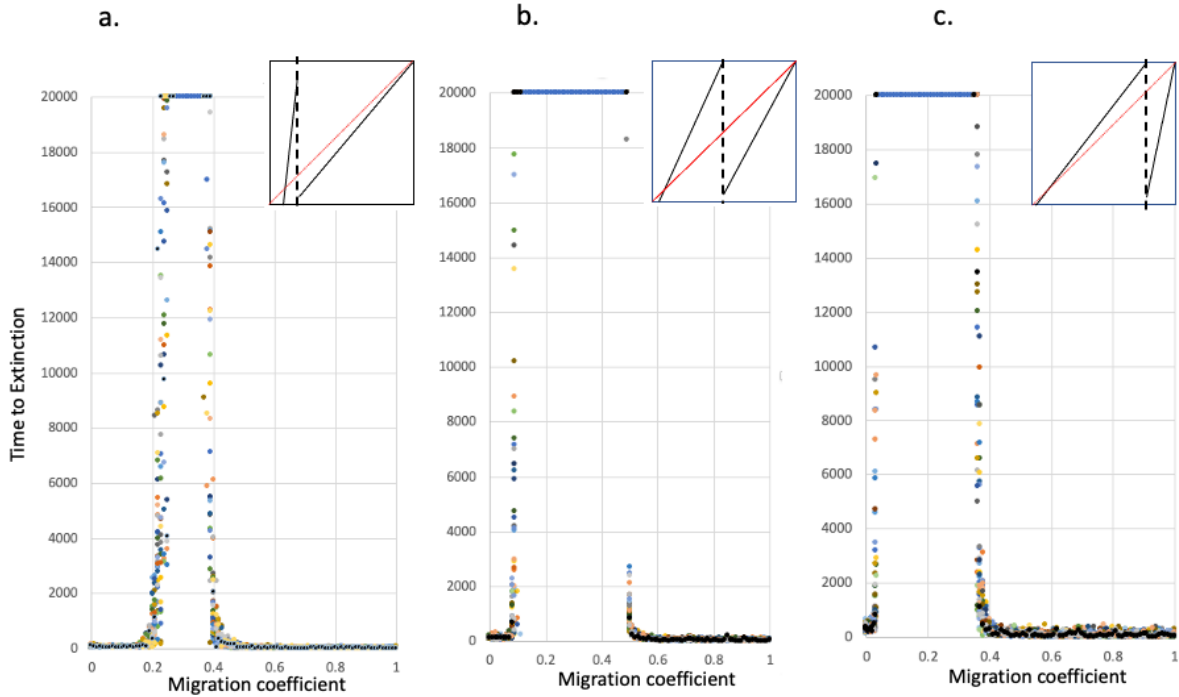

*Figure 18. Results of calculations for expected time to extinction for 20 subpopulations over a range of migration rates for three values of $x_0$. Extinction rate $(x_a – x_r) = 0.1 – 0.05 = 0.05$. a. $x_0 = 0.2$. b. $x_0 = 0.5$. c. $x_0 = 0.8$. Insets are graphs of the sawtooth map for each of the cases.*

     As discussed above, the left-hand border of the metapopulation window results from the combination of an accumulation of subpopulations into the near-zero symmetry group and the basin boundary collision of the migration-influenced chaotic attractor boundary and the Allee point. The right-hand boundary is generated in a completely different way. Synchrony is again involved, but its generation is distinct. Recalling the expected dynamics based on the derivative of the recursion function (equation 4 and Fig 10), we illustrate the process in figure 19. Based on the empirical fact and theoretical expectation that the model when including migration, has a derivative that determines an expansion of near points, in the classic stretching of a chaotic attractor versus the stretch-reversal of those points (see figure 10 and equations 9). Clearly, when the stretch expansion is stronger than the stretch-reversal, the model presents a chaotic trajectory. However, as the stretch-reversal component t becomes stronger, it acts to synchronize first individual subpopulations, and finally groups of subpopulations.

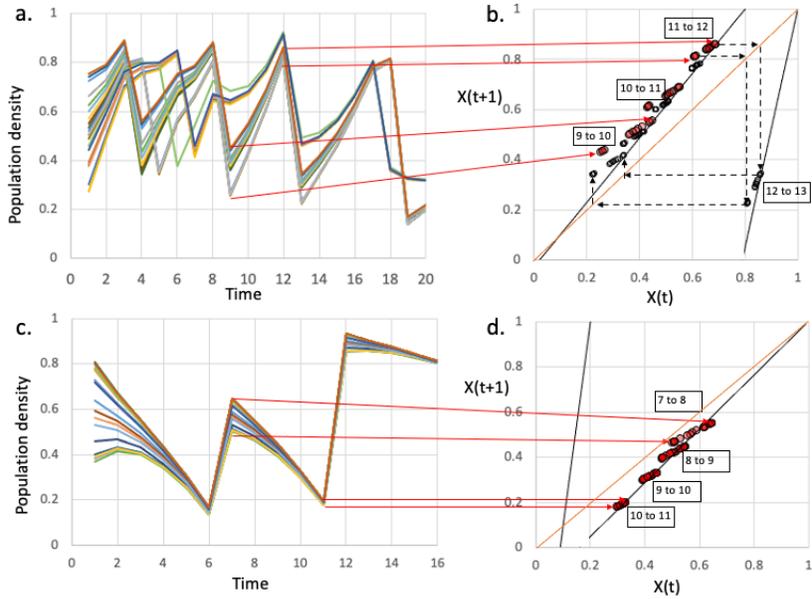

*Figure 19. Population trajectories (time series) and return maps for particular time points, illustrating stretch-reversal for two distinct values of $x_0$. a. twenty subpopulations with $x_0 = 0.8$, $m = 0.5$, $x_a = 0.1$, $x_r = 0.05$. b. return map for times 9 through 13 of the time series in part a. Note the stretch-reversal character of the model on the population increase part of the model. Red arrows connect the values in part a with the points on the return map and dashed black arrows indicate the dynamics on the return map. c. twenty subpopulations with $x_0 = 0.2$, $m = 0.5$, $x_a = 0.1$, $x_r = 0.05$. d. return map for times 7 through 11 of the time series in part c. Note the stretch-reversal character of the model, here on the population decline part of the model. Red arrows connect the values in part a with the points on the return map.*

The overall behavior of the model can be easily envisioned on the return map for a variety of parameter values, as displayed in figure 20, for $x_0 = 0.2$ (similar qualitative reasoning applies to all other values of $x_0$). With very low migration rates the basic operation of extinction through chaotic behavior transcending the Allee point creates the situation where adding migration to subpopulations that are destined to extinction (making them, formally, propagating sinks) does not create a metapopulation (Fig 20 a,b). However, there is a critical point at which the slope of the return map of the model (with migration) becomes less than 1.0 and, at the same time, the intercept at the parameter $x_0$ is above the Allee point, where the chaotic attractor is buffered away from that Allee point (effectively the structure of a basin boundary collision, where the Allee point is the separatrix of the basin that includes zero) to cancel all potential for extinction of any of the subpopulations (Fig 20 c, d, e). Further increase in migration generates a situation of stretch reversal (Fig 19), which increases the variability of the projections, reaching a critical point where the empirical model again intersects the Allee point and the permanent metapopulation is again lost through a basin/boundary collision (Fig 20 f,g).

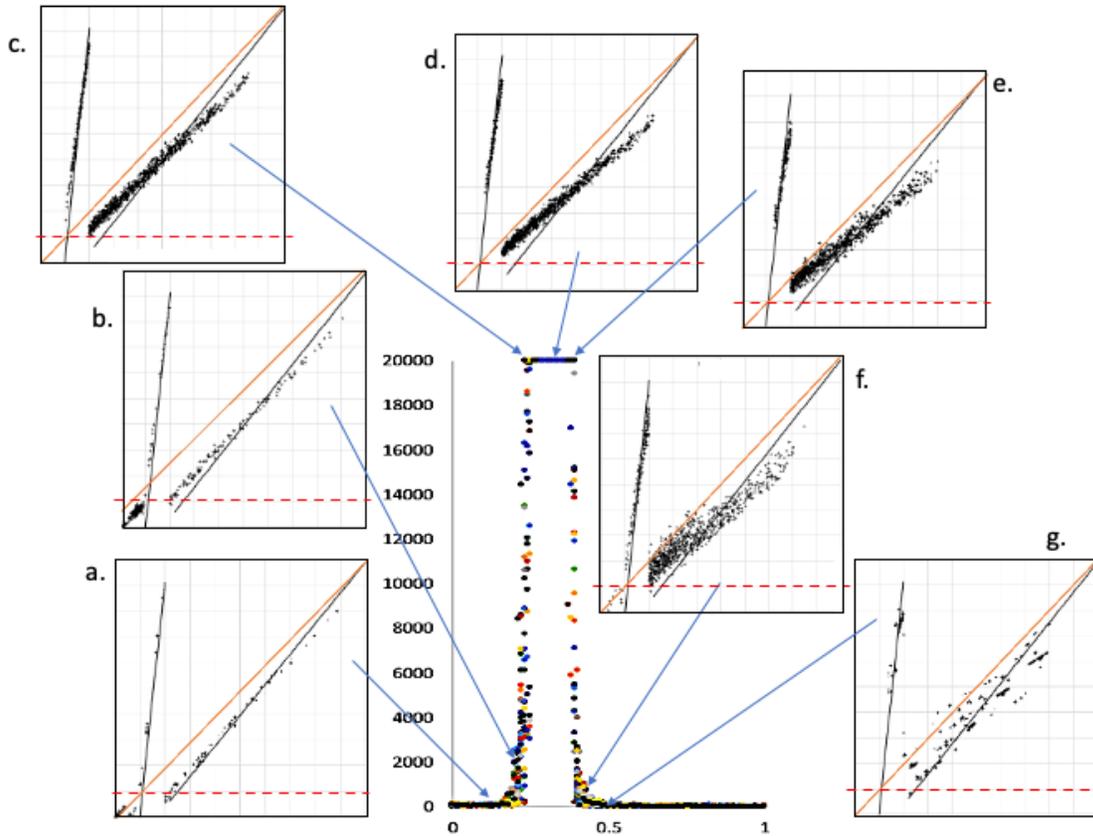

*Figure 20. The return map of the full model at various values of the migration coefficient. Central scattergram is the time to extinction as a function of migration coefficient (same as figure 18a), and each return map (a – g) includes the function (equations 7) as the black linear segments, the points from simulations of the model with migration (equations 8), with the Allee point illustrated with the horizontal red dashed line.*

The empirical size and position of the metapopulation window is shown in figure 21, effectively a summary of the ideas expressed in figure 18.

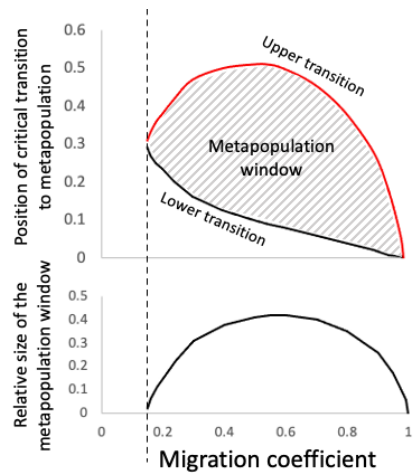

*Figure 21. Structure of the metapopulation window as a function of the migration coefficient.*

**4 Discussion.** The underlying assumptions of our iterative map approach are found frequently in the ecological literature. They are arguably essential for the deterministic functioning of a metapopulation. In particular there are three basic assumptions. We assume 1) an Allee point exists, 2) a chaotic-like attractor characterizes each subpopulation and 3) universal migration occurs (all subpopulations contribute a constant proportion of their populations to a general pool). Each of these three assumptions has a significant history in ecology.

The existence of an Allee point has been the subject of much recent theoretical study in the discrete time framework (e.g., Schreiber, 2003; Shabbir et al., 2020; 2020a; Din 2017; Vortkamp et al, 2020), reinforcing early ideas of complex dynamics emerging from evident ecological structures, such as predation. As a simplification, we have argued elsewhere that consideration of multiple trophic structures (e.g., a predator and pathogen simultaneously affecting a population with an Allee point), while obviously amenable to modeling in two or three dimensions (predator and prey and/or disease), could alternatively be envisioned in a simpler framing with an iterative map (Vandermeer and Perfecto, 2019; Vandermeer, 2021). As a common phenomenon in nature, considerable literature cites numerous empirical examples (e.g., Stephens and Sutherland, 1999; Levitan and McGovern, 2005) although detection in nature could be difficult due to sampling problems with rare populations.

Since Robert May (1974) reinterpreted the insights of Li and Yorke (1975), the subject of chaos has loomed large in the literature of theoretical ecology (Hastings et al., 1993; Rai and Schaffer, 2001; Ong and Vandermeer, 2015; Pearce et al., 2020). Generally, the style of theoretical work might be characterized as experimental mathematics since evidence for chaos is frequently presented in the form of complex time series and bifurcation diagrams (e.g., Vandermeer, 1993; 2004; 2006a) and only rarely visited with rigorous analytical tools (e.g., Drubi et al., 2021). Empirical evidence for chaotic dynamics has left little doubt that it is common (Cushing et al., 2001; Blasius et al., 1999; Benincà et al., 2008; 2009; but see also Berryman and Millstein, 1989). Also the idea of intermittent populations, closely related to chaotic ones, is relevant to this article, although there does not seem to be an extensive literature on the subject (Vandermeer, 2019; 2021). Finally, Schreiber (2003) put the issue of chaos and the Allee effect together in a way similar to the present work, using the Ricker model as a foundational framework. It goes without saying that the extinction modeled herein, as in previous studies, is dependent on a chaotic-like population that periodically descends below an Allee point.

Inter-population migration, the other element of basic metapopulation biology, has been a standard topic in ecology, especially after Levins' (1969) paper on metapopulations. In the context of metapopulation dynamics, migration takes on a variety of forms: for plants, seed dispersal; for viruses, inter-host transmission; for mammals, physical movement; for fish, swimming; for plankton, floating; birds, flying; spiders, ballooning; etc . . . The only point is that individuals in one subpopulation arrive at other subpopulations. One framing effectively cancels all structure out of the migration process and treats it almost as if it were nothing but simple diffusion (e.g., Pires et al., 2021). Although such an approach is obviously a caricature of any of the processes that fit into the migration term of a real metapopulation, it is a place to start, and is the general framework embraced in this work. All subpopulations contribute a constant fraction of their population density to a migratory pool and receive a constant fraction from that migratory pool (Fig. 1). Explorations of deviations from this key assumption are warranted, but not included in this study.

We employ two fundamental iterative map frameworks, the logistic/Allee map and the sawtooth map (a modification of the basic Bernoulli map). A curious and very general result emerges in both cases, that simply connecting propagating sinks with migration does not necessarily form a successful metapopulation. Indeed, under certain circumstances adding migration can not only fail to result in a metapopulation, it can hasten the time to extinction. This is especially true when population processes are dominated by negative intrinsic growth parameters. But it also emerges from subpopulation synchronization, as is well-known in the literature. We find two qualitatively distinct emergent mechanisms of synchronization either of which results in the failure of the metapopulation, or even in the metapopulation more rapidly extinguishing itself than happens with the completely unconnected subpopulations, contradictory to the intuitively expected outcome.

It is evident that the very essence of the metapopulation paradigm includes the process of local extinction. Consequently, we here emphasize the role of extinction in structuring the details of metapopulation behavior. First, modifying the classic logistic map with the addition of the Allee effect (Vandermeer and Perfecto, 2019), we explored various parameter states. With this more or less classic approach we find that the basic ideas of metapopulation are corroborated, but certain details emerge. One important result is the modifying nature of subpopulations synchronizing with one another. Although previous work clearly recognized this potential contradiction in metapopulation biology (e.g., Koelle and Vandermeer, 2005), we here explore the nature of that contradiction explicitly. Key to the overall behavior is the precise mechanism of synchronization, since that seems an important determinant of whether or not a group of propagating sinks will form a metapopulation. We find two distinct mechanisms whereby synchronization emerges – 1) through synchronous local extinction or 2) through what we term "stretch reversal."

First, subpopulation synchronization often occurs when two or more populations reach zero or below, wherein the underlying model resets to precisely zero. With a group of say 20 subpopulations (the basis of most simulations in this work), all of which are propagating sinks, induced to be so by intermittent chaos, the probability that two or more of them will reach zero or below is substantial, depending on parameters. Yet, because of the strictly deterministic nature of the model, such a simultaneous extinction means that both (or all) of those subpopulations will begin their resuscitated life with precisely the same number of individuals, and will consequently stay synchronized forever. Clearly this result will be modified if the migration terms are not constant. With some variability in the migratory inputs into the zero density populations, the strong synchronizing effect of local extinction will be moderated, the degree of moderation probably dependent on how strong either the pattern of, or stochastic effect on, migration might be. Consequently, it might be argued that this mechanism of synchrony-generation is an artifact of looking at the world in a strictly deterministic fashion, and perhaps that is true. But short of perfect symmetry, even approximate resetting every time a true extinction event happens, which must be the case since we are presuming the subpopulations are sink populations, will tend to synchronize the populations at least for a while. The underlying assumption of chaos will eventually cause a desynchronization as long as there is some variance in the fraction of the migration pool that resets those simultaneously extinct subpopulations, but this is itself an assumption of the deterministic nature of the model. Common sense suggests that the mechanism of simultaneous local extinction should, at least for some period of time, have the same effect in nature as it does with this model.

The second mechanism of synchrony is stretch reversal. Because both models are chaotic, they both result in the characteristic stretching and folding known to characterize chaotic attractors.

However, with sufficiently high migration rates, the process may be effectively reversed. As populations either increase or decrease exponentially, the migration process causes small populations to increase at a relatively larger rate than large ones, such that the underlying process causes the range of population densities to continually decline such that by the time they reach a high density and are shuttled off to the decreasing part of the map, they are relatively closely synchronized. It is not unusual that this type of behavior happens at very high migration rates, and is a reflection of the theoretical foundations that yield chaos in models such as the logistic map (see Fig 19).

Either during the process of metapopulation extinction, or in the arrival at metapopulation permanence, one feature of almost all of the parameter space is the formation of synchrony groups. As a mechanism of extinction, we have already noted this issue. However, as a phenomenon in and of itself it is an interesting issue, with subgroups forming rapidly and then declining in number as larger groups are formed from smaller ones (Figs 15, 17), reminiscent of other literature with a distinctly different theoretical framing (Hajian-Forooshani and Vandermeer, 2020; Vandermeer et al., 2021).

In both the Logistic/Allee and sawtooth models, the generalization that there is a balance between extinction and synchrony is, as expected, a strong result of the model. Details of its emergence are of interest, to be sure, but do not cancel the fundamental idea that the interaction of extinction and synchrony produce a generalized pattern in which a "metapopulation window" emerges over a range of migration coefficients (Figs 3, 18). The potential practical implications are evident for ecosystem management issues such as landscape biodiversity preservation, or regional planning for biological control in forestry or agriculture, or catch limitations in local fisheries management.

**Acknowledgements**: Prepared under support from NSF DEB 1853261, NIFA/USDA 2017-67019-26292, and NIFA/USDA 2018-67030-28239.